# Operando Insights on the Degradation Mechanisms of Rhenium-doped and Undoped Molybdenum Disulfide Nanocatalysts for Electrolyzer Applications

Raquel Aymerich-Armengol*, Miquel Vega-Paredes, Zhenbin Wang, Andrea M. Mingers, Luca Camuti, Jeeung Kim, Jeongwook Bae, Ilias Efthimiopoulos, Rajib Sahu, Filip Podjaski, Martin Rabe, Christina Scheu, Joohyun Lim* and Siyuan Zhang*

R. Aymerich-Armengol (rayar@dtu.dk), M. Vega-Paredes, A.M. Mingers, I. Efthimiopoulos, R. Sahu, M. Rabe, C. Scheu, S. Zhang (siyuan.zhang@mpie.de)
Max-Planck-Institut für Eisenforschung GmbH, Max-Planck-Straße 1, 40237, Düsseldorf, Germany

Z. Wang
Department of Materials Science and Engineering, City University of Hong Kong, Hong Kong SAR 999077, China

L. Camuti
Max-Planck-Institut für Festkörperforschung, Heisenbergstraße 1, 70569 Stuttgart, Germany
Department of Chemistry, University of Munich (LMU), Butenandtstraße 5–13 (D), 81377 Munich, Germany

J. Kim, J. Bae, J. Lim (jlim@kangwon.ac.kr)
Department of Chemistry, Institute for Molecular Science and Fusion Technology, Multidimensional Genomics Research Center, Kangwon National University, Chuncheon, Gangwon, 24341, Republic of Korea.

F. Podjaski
Department of Chemistry, Imperial College London, W12 0BZ London, United Kingdom

**Abstract**

MoS$_2$ nanostructures are promising catalysts for proton-exchange-membrane (PEM) electrolyzers to replace expensive noble metals. Their broadscale application demands high activity for the hydrogen evolution reaction (HER) as well as robust durability. Doping is commonly applied to enhance the HER activity of MoS$_2$-based nanocatalysts, but the effect of dopants in the electrochemical and structural stability is yet to be discussed. Herein, we correlate o*perando* electrochemical measurements to the structural evolution of the materials down to the nanometric scale by identical location electron microscopy and spectroscopy. The range of stable operation for MoS$_2$ nanocatalysts with and without rhenium doping is experimentally defined. The responsible degradation mechanisms at first electrolyte contact, open circuit stabilization and HER conditions are experimentally identified and confirmed with the calculated Pourbaix diagram of Re-doped MoS$_2$. Doping MoS$_2$-based nanocatalysts is validated as a promising strategy for the continuous improvement of high performance and durable PEM electrolyzers.

## 1. Introduction

Molybdenum disulfide (MoS$_2$) is widely regarded as a Pt alternative for the hydrogen evolution reaction (HER) in acidic media due to its high catalytic activity, abundance and low price[1-3]. However, it crystallizes in the thermodynamically stable 2H-MoS$_2$ polytype, which only possesses high HER activity on the edge sites[4]. In order to enhance the HER performance of MoS$_2$, exposure of the edge active sites can be maximized in nanostructures[5] e.g. with a nanoflower morphology[6, 7]. Other strategies to optimize the HER activity include sulfur-defect engineering[8, 9], phase engineering[10, 11], the use of amorphous MoS$_2$ layers[12] or doping with transition metals[13, 14].

In particular, Re-doped MoS$_2$ nanomaterials (Re$_x$Mo$_{1-x}$S$_2$) have shown remarkable improvement on the HER activity over MoS$_2$ nanocatalysts[15-18]. The n-type Re dopant not only acts as an electron donor, but also favors local transformation to 1T-MoS$_2$, while both effects contribute to improved HER performance[19-22]. Although the 1T-MoS$_2$ polytype provides enhanced conductivity and basal-plane HER active sites[11, 23], it is a metastable phase which can revert to the 2H phase. Nevertheless, Re-doping stabilizes the 1T phase and 1T-Re$_x$Mo$_{1-x}$S$_2$ has shown long lifetimes on the shelf and stability under HER conditions[15, 17]. In comparison, alkali metals can also stabilize 1T-MoS$_2$ nanocatalysts, but they are known to suffer from poor durability[22, 24].

Indeed, beyond the HER activity, the appeal of MoS$_2$-based nanocatalysts relies on the high stability during operation. However, such claim is usually only validated by electrochemical measurements with limited time (a few minutes to several hours) and without correlative characterization on the catalysts and electrolytes [25]. Only recently, through *operando* characterization using a scanning flow cell (SFC) connected to inductively coupled plasma mass spectrometry (ICPMS), bulk 2H-MoS$_2$[26] and [Mo$_3$S$_{13}$]$^{2-}$ clusters[27] have been confirmed as stable during galvanostatic holds in the cathodic HER regime in acidic conditions. Moreover, ICPMS measurements revealed high dissolution rates when the catalysts were subjected to open-circuit potential (OCP) conditions, a well-known problem for alkaline HER catalysts, which was only recently reported for acidic HER catalysts[28-31].

However, so far, the dissolution during OCP was only investigated for the initial minutes[26, 30], and its effects on the HER performance and the structure of the nanocatalysts remain unknown. Furthermore, the effect of dopants on the electrochemical and structural stability of MoS$_2$ (e.g. the highly active Re$_x$Mo$_{1-x}$S$_2$ nanocatalysts) and the degradation mechanisms still need to be clarified for both OCP and HER conditions.

Herein, we provide a systematic electrochemical stability study for Re$_x$Mo$_{1-x}$S$_2$ nanocatalysts by the combination of *ex-situ* electrochemical degradation and *operando* characterization performed by SFC-ICPMS. These measurements were correlated to the morphology and chemical composition evolution revealed by microscopy imaging and spectroscopy acquired in identical locations (IL). The results confirmed high electrochemical and structural stability of the materials during HER, quantitatively showing a higher stability of Re$_{0.2}$Mo$_{0.8}$S$_2$ nanocatalysts over MoS$_2$. Mechanisms of degradation during the first electrolyte contact, OCP stabilization and HER are proposed based on the experimental evidence from IL-scanning transmission electron microscopy (STEM), SFC-ICPMS, X-ray photoelectron spectroscopy (XPS) and X-ray absorption spectroscopy (XAS). According to the calculated Pourbaix diagrams, the main degradation mechanisms are based on the surface oxidation of Re$_x$Mo$_{1-x}$S$_2$ stemming from the synthesis and/or during OCP. These results highlight the excellent stability of pristine and Re-doped MoS$_2$-based nanostructures for use in acidic electrolyzers and establish a correlative methodology for the assessment of structural and electrochemical stability of doped MoS$_2$ nanocatalysts. The developed methodology and mechanistic insights will motivate the exploration of further metal dopants to simultaneously enhance the activity and stability of HER catalysts for broadscale commercial applications.

## 2. Results
### 2.1 Effect of rhenium in the activity and morphology of Re$_x$Mo$_{1-x}$S$_2$ nanocatalysts

Molybdenum disulfide nanocatalysts were grown on carbon paper (CP) substrates through a hydrothermal route[32] with various amounts of Re doping. Such substrate consists of interwoven carbon fibers with a diameter of ~10 μm. The stoichiometries of $Re_xMo_{1-x}S_2$ (x = 0, 0.2, 0.3 and 0.4) were determined by ICPMS. The HER activity of the electrodes was comparatively tested by cyclic voltammetry (CV) (**Figure S1**) in 0.5 M $H_2SO_4$ under continuous Ar purging from 0 to −0.25 $V_{RHE}$ before ohmic drop correction (iR correction). The results show that Re-containing materials have similar HER activity, with similar overpotential values at the same current density normalized to the metal (Mo and Re) content, which is significantly lower than the overpotential of bare $MoS_2$/CP electrode (**Figure 1a**). Considering the limited activity improvement from x = 0.2 to 0.4 and the scarcity of Re, the $Re_{0.2}Mo_{0.8}S_2$ composition was chosen for the stability study alongside $MoS_2$.

**Figure 1b** and **Table S1** present a comparison of the HER activity of the nanocatalysts in this work with literature reports of $MoS_2$ and $Re_xMo_{1-x}S_2$ nanocatalysts. It is evident that our electrocatalysts reach state-of-the-art performance both in terms of the overpotential at −10 mA/cm$^2$ ($\eta_{10}$) and the Tafel slope. Specifically, the $MoS_2$/CP and $Re_{0.2}Mo_{0.8}S_2$/CP Tafel slopes were 44 mV/dec and 55 mV/dec, respectively, while their $\eta_{10}$ values were 203 $mV_{RHE}$ and 136 $mV_{RHE}$ (**Figure 2a, b; Table S2**). Although these figures of merit evidence an improved $\eta_{10}$ for the nanocatalysts doped with Re, its Tafel slope is higher. The optimal rhenium composition to minimize the Tafel slope when compared to $MoS_2$ varied across reports [33, 34], which indicates additional factors contributing to the activity of $Re_xMo_{1-x}S_2$ nanocatalysts, e.g. the extend of 1T additional sites, sulfur vacancies or other defects.

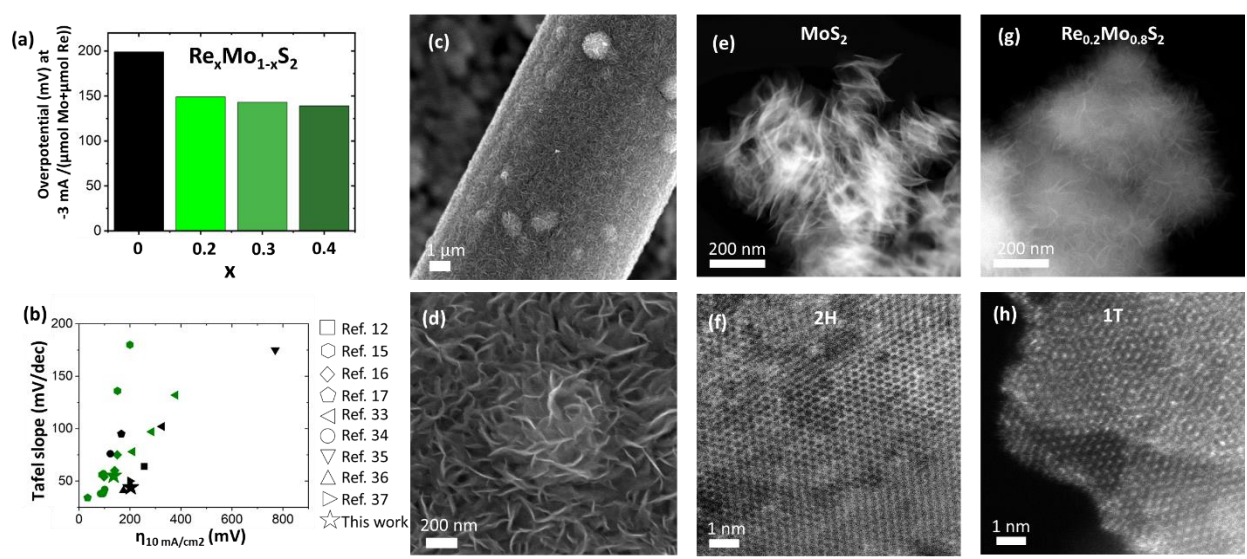

**Figure 1.** (a) Overpotential (normalized by loading) at a current density of −3 mA/(μmol Mo + Re) of $Re_xMo_{1-x}S_2$/CP materials with stoichiometries x=0, x=0.2, x=0.3 and x=0.4. (b)

Comparison of overpotential $\eta_{10}$ values and Tafel slopes from this work with literature references [12, 15-17, 33-37] of $MoS_2$ and $Re_xMo_{1-x}S_2$ catalysts. (c, d) SEM micrographs of $Re_xMo_{1-x}S_2$ nanocatalysts grown on a carbon paper. HAADF-STEM micrographs of (e) $MoS_2$ and (g) $Re_{0.2}Mo_{0.8}S_2$ nanocatalysts. High resolution HAADF-STEM images showing (f) 2H and (h) 1T phase areas present in $Re_{0.2}Mo_{0.8}S_2$ nanocatalysts.

The morphology of the $MoS_2$/CP and $Re_xMo_{1-x}S_2$/CP electrodes was compared with scanning electron microscopy (SEM) micrographs (**Figure 1c, d; Figure S2**). Both electrodes showed a layer of $Re_xMo_{1-x}S_2$ nanosheets covering the surface of the CP substrate. When synthesized as powders (in absence of the substrate), $Re_xMo_{1-x}S_2$ materials also exhibit a nanoflower morphology, as revealed by STEM (**Figure 1e, g**) and transmission electron microscopy (TEM) (**Figure S3**) micrographs. In addition, the packing density of the nanoflowers increases with increasing Re content. High resolution high angle annular dark field (HAADF)-STEM micrographs revealed the distribution of high-contrast Re within the $Re_{0.2}Mo_{0.8}S_2$ lattice (**Figure S4**). XPS indicates that Re is prevalent in the $Re^{4+}$ oxidation state showing binding energies that are typical for the disulfide form (**Figure S5**). Furthermore, as a result of Re incorporation, the phase transformation from the 2H to the 1T polymorphic structure was also observed in a few areas in $Re_{0.2}Mo_{0.8}S_2$ (**Figure 1 h**, **Figure S6**), whereas other areas remained 2H (see **Supplementary Note 1**, **Figure 1f**, **Figure S6**, **Figure S7**). The crystal structure of $MoS_2$ was consistently found to be only 2H (**Figure S7**, **Figure S8**).

## 2.2 Electrochemical and structural stability of $Re_xMo_{1-x}S_2$ nanomaterials

In order to make an accurate assessment of the stability of $Re_xMo_{1-x}S_2$ nanocatalysts, the contributions of different mechanisms of corrosion and dissolution need to be decoupled. Since recent reports[26, 30] suggest that OCP and electrolyte contact can severely degrade $MoS_2$-based nanocatalysts, the stability of the $Re_xMo_{1-x}S_2$ nanomaterials has been sequentially studied during early contact to the electrolyte, prolonged immersion at OCP conditions, and operation at HER potentials.

### 2.2.1 Stability at early electrolyte contact (immersion)

The dissolution and its kinetic evolution occurring to the fresh electrodes after synthesis was evaluated immersing them in 5 ml of electrolyte for 27 h, exchanging it every 9 h, and ICPMS analyses of Mo and Re dissolved in such electrolytes during 0-9 h, 9-18 h, 18-27 h are shown in **Figure S9a.**

The results reveal that both samples strongly dissolved during the first 9 h of contact with the electrolyte, much more than during the subsequent periods of 9 hours. Thus, we separately consider the dissolution during the initial 9 h as the effect of early electrolyte contact, while the subsequent hours would be attributed to other mechanisms derived from further stabilization at OCP. To narrow down the time span of the early electrolyte contact with significant dissolution, further sets of measurements were conducted on the dissolution after every hour for the initial 6 hours (**Figure S10**). The results show that for both $MoS_2$ and $Re_{0.2}Mo_{0.8}S_2$ nanocatalysts, the majority (~90% of the total) dissolution had happened during the first hour.

To correlate the dissolution results to the structural evolution and find the responsible mechanisms, the electrodes were investigated by means of IL-SEM (**Figure 2 a-d**) before and after the electrolyte immersion for 27 h. For both $Re_xMo_{1-x}S_2$/CP electrodes, small differences were spotted in some locations in the form of detached catalyst particles from the substrate (see arrows in the figures). The loss of catalysts can be correlated to the dissolution measured by *ex-situ* ICPMS, which predominantly occurred during the first hour (**Figure S9a, Figure S10**). On the other hand, the thickness of the remaining layer remained stable, from an average of 556 ± 20 nm before immersion to 553 ± 13 nm after immersion for $MoS_2$/CP and from 451 ± 20 nm to 452 ±10 nm for $Re_{0.2}Mo_{0.8}S_2$/CP (**Figure S11**, **Figure S12**, **Figure S13**).

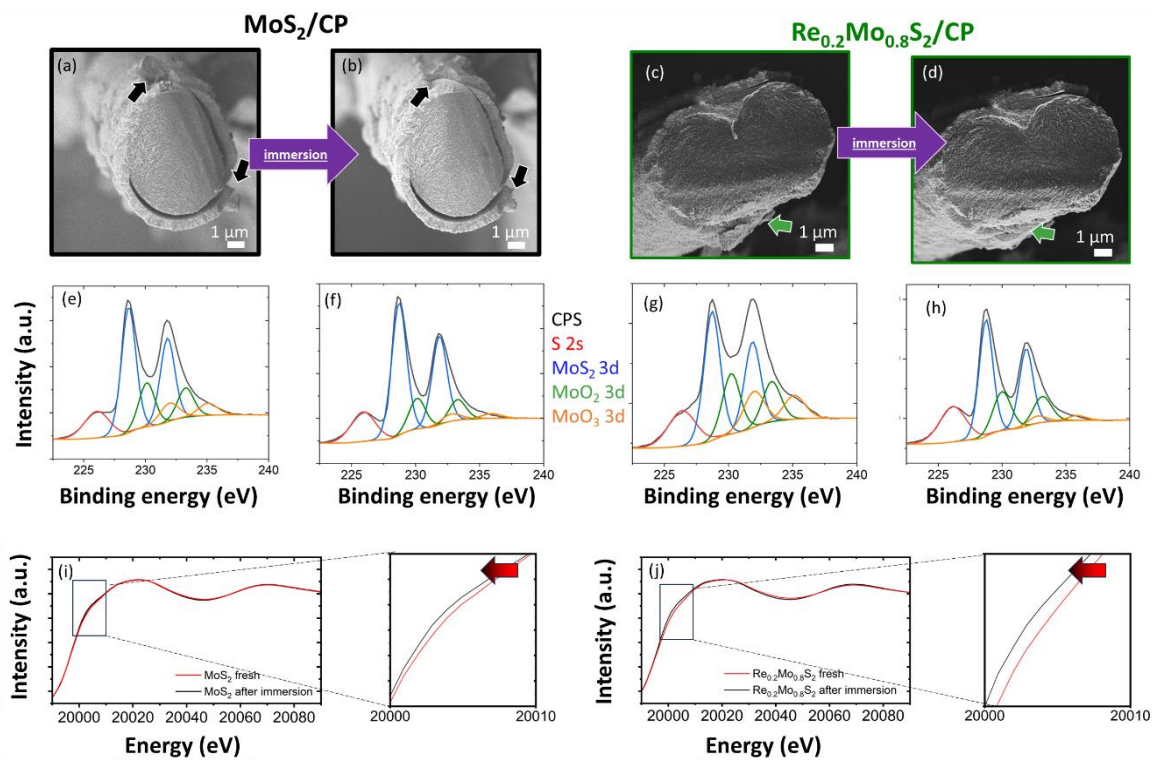

**Figure 2.** SEM micrographs before and after immersion in 0.5 M $H_2SO_4$ for 27 h for (a, b) $MoS_2$/CP and (c, d) $Re_{0.2}Mo_{0.8}S_2$/CP, with black and green arrows highlighting detached catalyst particles. XPS spectra of Mo 3d showing the contributions of S, $MoS_2$, $MoO_2$ and $MoO_3$ peaks before (e, g) and after (f, h) 27 h of immersion for $MoS_2$/CP and $Re_{0.2}Mo_{0.8}S_2$/CP electrodes. XAS spectra of Mo K-edge of (i) $MoS_2$/CP and (j) $Re_{0.2}Mo_{0.8}S_2$/CP electrodes before and after 27 h of immersion.

Additionally, XPS was conducted on the electrodes to understand the changes in the surface composition of the electrodes. **Figure 2 e-h** show the deconvoluted Mo 3d spectra for both samples before and after immersion for 27 h. The results reveal the presence of the oxide species $MoO_2$ and $MoO_3$ on both electrodes as-synthesized before immersion (see **Figure S14** for more details on the peak assignment). These oxides arise from the incomplete reaction of the molybdenum precursor reported to happen at the synthesis temperature of 200 °C, which was also correlated to enhanced HER performance [38, 39]. The presence of $Mo^{6+}$ oxides is strongly reduced after immersion for both electrodes (**Figure S5**, **Figure S14**, **Figure S15**), indicating that the strong Mo dissolution observed during the initial hours of OCP (**Figure S9a, Figure S10**) is indeed related to partially unreacted reagents from the synthesis and/or storage of the material. This mechanism could also explain the over-stoichiometric dissolution of Mo from $Re_{0.2}Mo_{0.8}S_2$ compared to rhenium during the initial electrolyte contact. According to XPS, $Re^{4+}$ species (same oxidation state as $ReS_2$) are predominant on the $Re_{0.2}Mo_{0.8}S_2$ surface, which was maintained very stable after immersion (**Figure S9b, Figure S5**). To further confirm these results, XAS was conducted on both the $MoS_2$/CP and $Re_{0.2}Mo_{0.8}S_2$/CP electrodes before and after 27 h immersion in 0.5 M $H_2SO_4$. The Mo K-edges (**Figure 2i, Figure 2j**) of both electrodes show a shift towards lower energies after immersion, suggesting reduction of Mo species, which corroborates with the dissolution of $MoO_3$ species observed by XPS. Moreover, the Mo-K edge of $Re_{0.2}Mo_{0.8}S_2$/CP electrode onsets at lower energy than $MoS_2$/CP electrode (**Figure S16a**), which can be attributed to the coupling effect of the n-type Re-dopant on the electronic structure of the Mo atoms[40]. Finally, the Re $L_3$-edge XANES (**Figure S16b**) of $Re_{0.2}Mo_{0.8}S_2$/CP electrode does not show an obvious chemical shift after 27 h of immersion, again corroborating the XPS observation.

### 2.2.2 Stability at OCP conditions

As observed in **Figure S9a** and **Figure S10**, after the initial contact (0-9 h or first hour) to the electrolyte, the dissolution of the $Re_xMo_{1-x}S_2$/CP electrodes approaches a much lower rate, which needs to be explained by other degradation mechanisms at stabilized OCP conditions.

To understand the long-term effect of the OCP dissolution on the HER performance of the materials, CV measurements were performed with the $Re_xMo_{1-x}S_2$/CP electrodes after 13 and 30 days of OCP electrolyte immersion in ~5 ml of volume (**Figure S17**). Given the small volume of electrolyte, minimal changes of 2 mV of overpotential in both materials after the 30 day treatment were observed, which is within the experimental error. These results point to the build-up of dissolved species in solution preventing further degradation of the electrodes[41, 42], and highlights the importance of the operando measurements in the flow cell where the electrolyte is continuously refreshed to properly understand the intrinsic stability of the electrocatalysts.

Thus, to gain insights into such OCP degradation, *operando* electrochemical and dissolution measurements were performed with SFC-ICPMS[43]. **Figure 3** shows the time-resolved dissolution of Mo and Re in $MoS_2$ and $Re_{0.2}Mo_{0.8}S_2$ as a function of the potential applied in steps of 0.1 V from the HER regime $-0.1$ $V_{RHE}$ to 0.5 $V_{RHE}$, which is close to the OCP of both catalysts, as shown in **Figure S18a**. The dissolution observed during HER potential ($-0.1$ $V_{RHE}$) was negligible for all catalysts. When transitioning to more anodic potentials towards the OCP (0.5 $V_{RHE}$), Mo atoms start to dissolve, which aligns with the ex-situ ICPMS dissolution trends. Previous *ex-situ* experimental reports suggested dissolution at 0.23 $V_{RHE}$[30], while our *operando* data provides evidence for Mo dissolution starting at potentials as low as 0 $V_{RHE}$. In comparison, Re dissolution starts at more anodic potentials, becoming noticeable first at 0.3 $V_{RHE}$ and more pronounced towards the OCP, 0.5 $V_{RHE}$.

More severe dissolution at OCP conditions with respect to HER was further confirmed by *operando* measurements under other electrochemical protocols, including CV and chronopotentiometry measurements in the galvanostatic mode (**Figure S19**). Longer measurements maintaining OCP for ~1 h revealed that Mo dissolution starts to decay after an initial increase, while the Re dissolution slightly increased with time (**Figure S20**). Such result is in agreement with the *ex-situ* ICPMS measurements of the electrolyte after every 9 hours at OCP, further indicating that with more time stabilized at OCP, Mo and Re dissolution would approach to the stoichiometric ratio (**Figure S9b**).

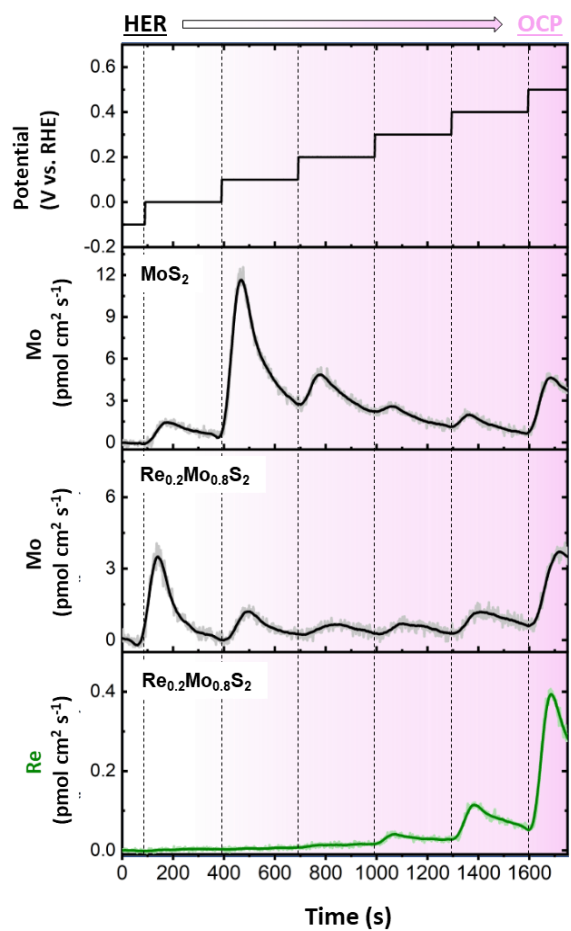

**Figure 3.** *Operando* SFC-ICPMS data showing the Mo and Re dissolution transitioning from HER potential (−0.1 $V_{RHE}$) to the OCP (0.5 $V_{RHE}$) for $MoS_2$ and $Re_{0.2}Mo_{0.8}S_2$ nanocatalysts.

To investigate long-term effects of OCP on the chemical composition and structure of the $Re_xMo_{1-x}S_2$ nanocatalysts, IL-STEM measurements were conducted before and after 13 days immersion in 0.5 M $H_2SO_4$ (**Figure 4**). The electrochemical measurements are performed directly on Au/C TEM finder grids that are numerated to allow the analysis of the same area of the specimen before and after electrochemical treatment, which provides information on the mechanisms of morphological degradation of the catalysts[44-46]. The OCP of $Re_{0.2}Mo_{0.8}S_2$ nanocatalysts dispersed on the Au/C TEM finder grid in the identical location set-up (**Figure S21**) was determined to be ~0.6 $V_{RHE}$, which is comparable to the OCP of $Re_xMo_{1-x}S_2$/CP electrodes (**Figure S18b**).

As shown in **Figure 4a-d,** there are no apparent changes in the morphology of $MoS_2$ and $Re_{0.2}Mo_{0.8}S_2$ nanocatalysts after 13 days of OCP beyond the small movement of the nanoflower layers (**Figure S22**, **Figure S23**). By tracking identical locations of the nanocatalysts, systematic changes in the chemical composition were detected by STEM-energy dispersive X-ray spectroscopy (EDS). The Re content in $Re_{0.2}Mo_{0.8}S_2$ increased from an average of 21.3 at% (normalized to Re+Mo) to 23.4 at% after 13 days of OCP (**Table S3**, **Figure S24**). This is consistent with the initially higher amount of Mo dissolution at OCP compared to Re dissolution (**Figure S9**).

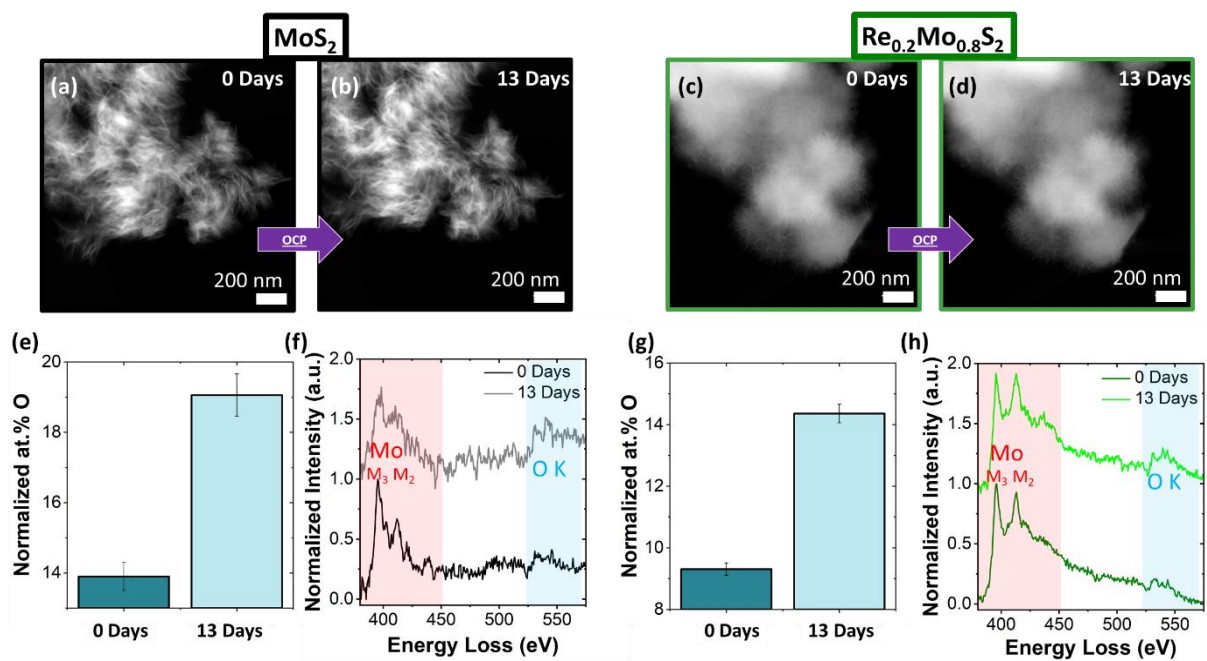

**Figure 4.** Morphology of (a, b) $MoS_2$ and (c, d) $Re_{0.2}Mo_{0.8}S_2$ nanoflowers before and after 13 days of immersion in 0.5 M $H_2SO_4$. Evolution of the O content $MoS_2$ and $Re_{0.2}Mo_{0.8}S_2$ nanocatalysts by EDS (e, g) and EELS (f, h) before and after 13 days immersion.

EDS measurements were also conducted at identical locations before and after 13 days of immersion to monitor the oxygen content of the nanocatalysts (**Table S3**). The results show an increase in the O content from 10.1 at% (normalized to O + S) to 12.6 at% for $MoS_2$ (**Figure 4e**, **Figure S25a, b**) and 6.6 at% to 7.9 at% for $Re_{0.2}Mo_{0.8}S_2$ (**Figure 4g**, **Figure S25c, d**). Oxidation of both samples after long term immersion is further verified by conducting electron energy loss spectroscopy (EELS) at identical locations, as shown in **Figure 4f**, **h**. For both catalysts, the intensity of the O-K edge increased after 13 days of immersion. Moreover, the oxidation state of Mo also increased after immersion. The intensity ratio of the Mo $M_{2,3}$ white lines ($I_{M3}/I_{M2}$) is lower in more oxidized states, as was demonstrated for $MoO_2$ and $MoO_3$ standards[47]. The $I_{M3}/I_{M2}$ decreased from 1.76 to 1.45 for $MoS_2$ and from 2.29 to 1.97 for $Re_{0.2}Mo_{0.8}S_2$, confirming an increase in the oxidation state of Mo in both nanocatalysts after 13 days immersion. Although $MoS_2$ and $MoO_2$ both have an oxidation state of $Mo^{4+}$, the higher electronegativity of oxygen compared to sulfur could also account for the decreased $I_{M3}/I_{M2}$ ratio.

**2.2.3 Stability at HER conditions**

To study the electrochemical stability of MoS$_2$/CP and Re$_{0.2}$Mo$_{0.8}$S$_2$/CP, the electrodes were immersed in 0.5 M H$_2$SO$_4$ for 32 h prior to HER conditions to completely remove the oxides from synthesis/storage, which would cause severe dissolution (see **Section 2.2.1**). After the treatment, 4000 CV cycles were subsequently performed in a fresh batch of electrolyte (0.5 M H$_2$SO$_4$) free of the dissolution products from the previous immersion step. Most CV cycles were conducted at a scan rate of 100 mV/s, and slower cycles at a scan rate of 1 mV/s were performed every 1000 CV cycles. For each cycle, the potential was scanned between 0 and −0.25 V$_{RHE}$ (before iR correction), reaching a current density of −16 mA/cm$^2$ for MoS$_2$/CP and −35 mA/cm$^2$ for Re$_{0.2}$Mo$_{0.8}$S$_2$/CP. The electrodes showed remarkable HER performance stability during the CV cycles, with negligible changes in η$_{10}$ (2 mV and 0 mV for MoS$_2$/CP and Re$_{0.2}$Mo$_{0.8}$S$_2$/CP, respectively), which are within the error of the analysis (**Figure 5a**, **Figure S26**, **Table S2**). Moreover, the changes in Tafel slopes are also minor (6 mV/dec and 3 mV/dec for MoS$_2$/CP and Re$_{0.2}$Mo$_{0.8}$S$_2$/CP, respectively). The HER stability demonstrated by the Re$_{0.2}$Mo$_{0.8}$S$_2$/CP electrodes is in agreement with previous reports: For example, η$_{10}$ of Re$_{0.55}$Mo$_{0.45}$S$_2$ nanocatalysts only increased by 2 mV after 3000 CV cycles[15].

Additionally, the dissolution of the electrodes during these 4000 CV cycles was quantified by *ex-situ* ICPMS (**Figure S9a**). Compared to the integrated dissolution during the ~8 h (4000 CV cycles) at HER conditions, the dissolution at OCP conditions is higher, which is consistent with the operando SFC-ICPMS measurements (**Section 2.2.2**). However, it is important to note that such *ex-situ* measurements possess limitations that hinder a full quantitative analysis. Firstly, although fresh electrolytes free of dissolved Mo and Re were exchanged prior to the CV cycles at HER conditions, *ex-situ* experimental protocols cannot avoid some time (at least of the order of minutes) when the electrodes are immersed in conditions closer to OCP. Secondly, while the previous immersion step could remove most oxides formed during the synthesis or storage, it would also introduce new oxides when approaching OCP (**Section 2.2.2**). These two issues can cause overestimation of the dissolution related to HER due to the degradation of oxides formed on MoS$_2$. Thirdly, the possible buildup of dissolved species in the stagnant electrolyte and the presence of a Nafion membrane in the H-cell can cause underestimation of the dissolution [41, 42]. As a result of these three considerations, the integrated dissolution during the CV cycles would lead to an inaccurate HER stability assessment.

To overcome the limit of *ex-situ* measurements, the stability of $MoS_2$ and $Re_{0.2}Mo_{0.8}S_2$ nanocatalysts under different HER potentials was investigated using the *operando* SFC-ICPMS set-up. $MoS_2$ and $Re_{0.2}Mo_{0.8}S_2$ nanocatalysts were drop cast onto FTO substrates and rinsed with 0.1 M $H_2SO_4$ prior to SFC-ICPMS measurement. The rinsing step significantly reduced the dissolution derived from the initial flowing electrolyte (0.1 M $H_2SO_4$) contact through the SFC. Once the dissolution signal decreased below the background level, the electrodes were subjected to HER conditions by chronoamperometry measurements at −0.05, −0.1, −0.2, −0.3, and −0.4 $V_{RHE}$. As shown in **Figure 5b,** there is negligible dissolution from $MoS_2$ and $Re_{0.2}Mo_{0.8}S_2$ between −0.05 and −0.3 $V_{RHE}$. Starting from −0.4 $V_{RHE}$, dissolution of Mo was observed from both $MoS_2$ and $Re_{0.2}Mo_{0.8}S_2$, while Re remained stable against dissolution. Thus, these *operando* measurements confirm the HER stability of both $MoS_2$ and $Re_{0.2}Mo_{0.8}S_2$, as long as the applied potential does not exceed −0.3 $V_{RHE}$, which is already a high overpotential for HER operation.

To specifically compare the stability of $MoS_2$ and $Re_{0.2}Mo_{0.8}S_2$ electrodes, the universal stability metric S*tability number* (S-number[48]) was used, which is a dimensionless figure defined as the ratio between the molar quantity of reaction products ($H_2$ for HER [30]) and the molar quantity of dissolved ions from the electrode (See **Supplementary Note 2**). At −0.3 $V_{RHE}$, within the stable range of HER operation, the S-number is evaluated to be at least $10^6$ for both $MoS_2$ and $Re_{0.2}Mo_{0.8}S_2$. This number is comparable to noble $IrO_2$ catalysts for acidic oxygen evolution reaction ($10^4 \sim 10^7$)[48], and way above photocatalysts such as $BiVO_4$ ($10^2$)[49]. At −0.4 $V_{RHE}$, where the Mo dissolution is clearly above background, the S-number of $Re_{0.2}Mo_{0.8}S_2$ ($4.8 \cdot 10^5$) shows a higher stability than bare $MoS_2$ ($3.4 \cdot 10^5$). Notably, Re dissolution was not observed even at −0.4 $V_{RHE}$, indicating a higher stability of Re than Mo in the HER regime. These quantitative results confirm that in addition to providing higher HER activity, Re-doping is a valid strategy to improve the electrochemical stability of $MoS_2$ nanocatalysts.

Finally, the electrochemical results were correlated with the morphology evolution of $MoS_2$/CP and $Re_{0.2}Mo_{0.8}S_2$/CP electrodes before and after 4000 CV cycles by means of IL-SEM. No visible changes were observed by comparing the SEM micrographs acquired at the same CP fibers, as the coverage and thickness of the nanocatalyst layer remained constant (**Figure S27, Figure S11**). The average thickness of such layer changed from 553 ± 13 nm to 555 ± 18 nm after HER for $MoS_2$/CP, and 452 ± 10 nm to 452 ± 10 nm for $Re_{0.2}Mo_{0.8}S_2$/CP, confirming that negligible dissolution took place. In addition, our recent report of IL-STEM on $Re_{0.2}Mo_{0.8}S_2$ nanocatalysts confirmed that the morphology and the composition were maintained down to the nanometric scale after 4000 CV cycles in the HER regime [46].

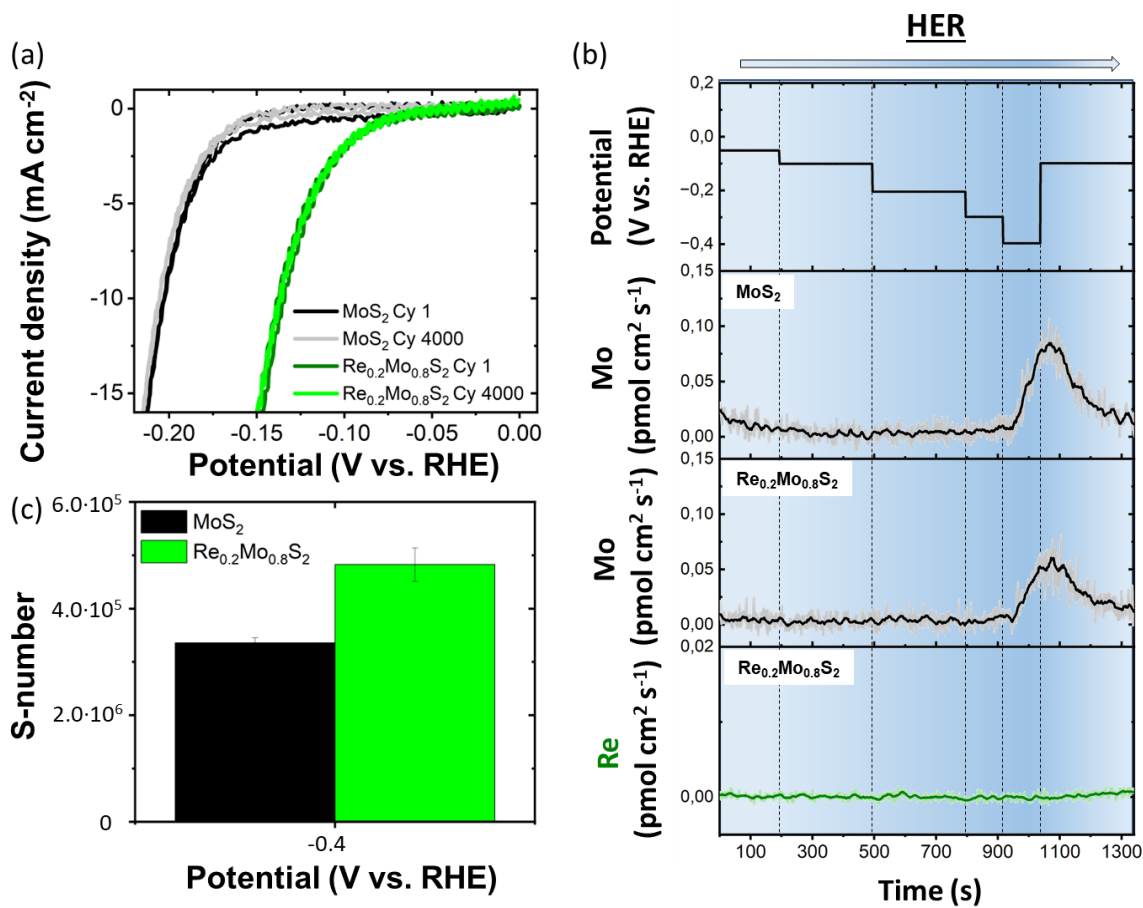

**Figure 5.** (a) CVs comparing the HER performance of $MoS_2$/CP and $Re_{0.2}Mo_{0.8}S_2$/CP electrodes at cycle 1 and cycle 4000. (b) *Operando* SFC-ICPMS data showing the Mo and Re dissolution at different HER potentials. (c) S-number of $MoS_2$ and $Re_{0.2}Mo_{0.8}S_2$ at −0.4 $V_{RHE}$.

### 3. Discussion: Mechanisms of dissolution

Considering the results from the electrochemical and structural stability studies, the following mechanisms of degradation of $Re_xMo_{1-x}S_2$ nanocatalysts can be established. We also separate the discussion into the three stages, consisting of the initial contact with the electrolyte, stabilization at OCP and the HER conditions.

The initial electrolyte contact causes most of the dissolution, especially severe in the case of Mo. This can be attributed to the dissolution and physical detachment of $MoO_3$ and $MoO_2$ stemming from the synthesis and storage (**Figure 2**, **Figure S9**, **Figure S14**). After most of the oxides are dissolved at the initial contact with the electrolyte, the OCP stabilizes at ~0.5 $V_{RHE}$, leading to a steady rate of Mo and Re dissolution.

To understand the origins of such OCP degradation, we conducted Pourbaix analysis calculating the diagram of $Re_{0.2}Mo_{0.8}S_2$ (**Figure S28**). **Figure 6** presents the calculated aqueous decomposition free energy ($\Delta G_{pbx}$) for $MoS_2$ and $Re_{0.2}Mo_{0.8}S_2$ as a function of the applied potentials. A larger $\Delta G_{pbx}$ indicates greater instability of the given species. These calculations show that the sulfide species remain stable at anodic potentials until $MoS_2$ transforms into $MoO_2$ at ~0.35 $V_{RHE}$, while $ReS_2$ is maintained stable up until ~0.4 $V_{RHE}$. These findings correspond with our experimental results obtained through *operando* SFC-ICPMS, where the onset of Re dissolution is delayed relative to that of Mo when approaching OCP potentials (**Figure 3**). Furthermore, the dissolution rate of both elements reach a stoichiometric ratio once the OCP is stabilized, as experimentally shown in **Figure S9**, **Figure S20**.

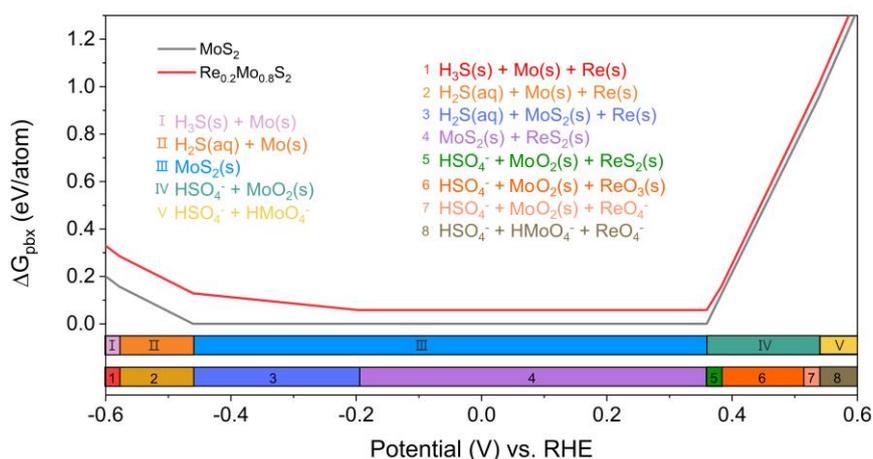

**Figure 6**. Calculated Pourbaix diagram for $MoS_2$ and $Re_{0.2}Mo_{0.8}S_2$ including their calculated aqueous decomposition free energy ($\Delta G_{pbx}$) as grey and red lines, respectively, with aqueous ion concentrations of $10^{-6}$ M at pH 1 and 25 °C. The projection of $\Delta G_{pbx}$ onto the potential axis represents the stable species within the corresponding regions on the Pourbaix diagram. Numerals index the relevant decomposition products of the potential region. Note that the small difference in $\Delta G_{pbx}$ (0.058 eV/atom) between $Re_{0.2}Mo_{0.8}S_2$ and $MoS_2$ in region 4 is attributed to the metastable structure employed in the calculations.

When subjecting the $Re_xMo_{1-x}S_2$ electrodes to cathodic HER conditions, Pourbaix analysis predicts a wide stability range for Mo at pH 1 from until ~−0.45 $V_{RHE}$, where a transformation from $MoS_2$ to Mo is predicted to occur (**Figure 6**). This is further confirmed experimentally by SFC-ICPMS, although the stability range is narrowed down to −0.3 to 0 $V_{RHE}$ (**Figure 5b**).

In the case of Re, experimentally no dissolution was observed *operando* at HER potentials as low as −0.4 $V_{RHE}$ (**Figure 5b**), despite the Pourbaix diagram predicting the transformation from $ReS_2$ to $Re^0$ at ~ −0.2 $V_{RHE}$. Given that no dissolution was detected with the high-sensitive SFC-ICPMS technique, we hypothesize that a higher overpotential would be required to drive such reaction, kinetically stabilizing the $Re_xMo_{1-x}S_2$ material. Note that $Re^+$ and $Re^-$ species were not considered for the $Re_xMo_{1-x}S_2$ Pourbaix diagram calculations based on experimental considerations, see **Supplementary Note 3, Figure S29** for further details. Considering the operando results, we note that the degradation observed at HER conditions when performing ex-situ analysis of the electrolyte after 4000 CVs from 0 to –0.25 $V_{RHE}$ (prior to iR correction) (**Figure 5a**, **Figure S9a**), can be attributed to the degradation of the $MoO_x$ produced during OCP and/or remaining from synthesis instead of an inherent instability of the $MoS_2$. Specifically, according to the Mo Pourbaix diagram, $MoO_2$ dissolves as $Mo^{3+}$ at pH 1 at potentials negative than 0.1 $V_{RHE}$ [50]. This highlights the need for pretreatment of the electrodes to eliminate the present oxides that can lead to misleading stability assessments. Based on such mechanisms, it also becomes evident that start-stop cycles intermixing OCP conditions ("stop", sulfide species transform into oxides) and HER operation ("start", oxides dissolve) can decrease the lifetime of $MoS_2$-based nanocatalysts for electrolyzers and should consequently be avoided. With these precautions, $MoS_2$ and $Re_{0.2}Mo_{0.8}S_2$ can be operated at a high stability for hydrogen production in the wide range of potentials from −0.05 to −0.3 $V_{RHE}$, with the Re-doped material surpassing the $MoS_2$ in both activity and stability.

**Conclusions**

The electrochemical and structural stability of $MoS_2$ and $Re_xMo_{1-x}S_2$ nanocatalysts was investigated through a combination of *operando* electrochemical measurements and correlative electron microscopy analyses at identical locations. We developed reliable quantification to demonstrate outstanding stability of $MoS_2$ during HER, and even superior HER activity and stability of $Re_xMo_{1-x}S_2$.

Based on the electrochemical stability and the structural evolution tracked by IL-SEM, IL-STEM, XPS, and XAS, we propose different degradation mechanisms of $Re_xMo_{1-x}S_2$ nanocatalysts at first electrolyte contact, OCP stabilization and HER conditions. Most of Mo and Re dissolution was measured upon first contact with the electrolyte, derived from a heavy loss of molybdenum oxide species ($MoO_2$, $MoO_3$) introduced from synthesis and storage. OCP stabilization led to the surface oxidation of $Re_xMo_{1-x}S_2$, resulting in their dissolution. Finally, HER conditions led to the dissolution of oxides formed during OCP stabilization, whereas the dissolution of sulfides remained negligible until −0.4 $V_{RHE}$ when Mo started dissolving. Thus, by eliminating most of the surface oxides from $Re_xMo_{1-x}S_2$, and using operando SFC-ICPMS, we could demonstrate their excellent stability during HER between −0.05 and −0.3 $V_{RHE}$, exceeding the stability number of $10^6$. Overall, the Re-doped $MoS_2$ nanocatalyst showed higher stability due to the lack of Re oxides created from synthesis and storage, the higher onset of oxidation of $ReS_2$ during OCP and the lack of dissolution at HER regime experimentally demonstrated until −0.4 $V_{RHE}$.

These results describe the degradation mechanisms of $Re_xMo_{1-x}S_2$ and explain the superior HER activity and stability of Re-doped $MoS_2$ catalysts when compared to bare $MoS_2$, opening the door to the rational design of doped-$MoS_2$ for optimized stability of PEM electrolyzers.

**Experimental section**

**Materials**

The synthesis of $Re_xMo_{1-x}S_2$ nanoflowers was conducted using the reagents ammonium heptamolybdate tetrahydrate (($NH_4)_6Mo_7O_{24}·4H_2O$, 99.98%, Sigma-Aldrich), ammonium perrhenate ($NH_4ReO_4$, >99%, Merck), thiourea ($SC(NH_2)_2$, >99.0%, Sigma-Aldrich) and ethanol (EtOH, >99.8 %, Carl Roth). Hydrophilic carbon paper (CP, HCP030N, Hesen) was employed as electrode substrate for the growth of $Re_xMo_{1-x}S_2$ nanocatalysts. The electrolyte of choice for all electrochemical measurements was diluted sulfuric acid ($H_2SO_4$, Suprapur, Sigma-Aldrich and MilliQ water).

**Synthesis of $Re_xMo_{1-x}S_2$ nanocatalysts and $Re_xMo_{1-x}S_2$/CP electrodes**

In a typical synthesis, 35.3 mg of $(NH_4)_6Mo_7O_{24}·4H_2O$ and 80.7 mg of $SC(NH_2)_2$ were dissolved in 10 ml of distilled water, adding 13.4 mg, 45.3 mg or 134 mg of $NH_4ReO_4$ for $Re_{0.2}Mo_{0.8}S_2$, $Re_{0.3}Mo_{0.7}S_2$ and $Re_{0.4}Mo_{0.6}S_2$ materials, respectively. The clear solution was transferred to a Teflon liner including a 1 cm × 2 cm CP with an area of 1 cm × 1 cm covered with Teflon tape, thus limiting the $Re_xMo_{1-x}S_2$ deposition to the 1 $cm^2$ remaining exposed. The solution with the substrate was heated at 200°C for 20 h in an autoclave and subsequently let to cool down at room temperature before removing the Teflon tape and rinsing the $Re_xMo_{1-x}S_2$/CP electrodes with water. Independently of Re content, the resulting electrodes showed a characteristic black color in the area with deposited material. If no substrate was introduced inside the Teflon liner and the concentration of all reagents was increased 2.7·$10^3$ times, black suspensions of $Re_xMo_{1-x}S_2$ nanocatalysts were obtained as a powder to analyze in ILSTEM. These products were washed by centrifuge-redispersion cycles in water and ethanol before evaporating the ethanol solvent at 110°C for 4 h. Black $Re_xMo_{1-x}S_2$ powders were subsequently ground in an agate mortar. The $MoS_2$, $Re_{0.2}Mo_{0.8}S_2$, $Re_{0.3}Mo_{0.7}S_2$ and $Re_{0.4}Mo_{0.6}S_2$ stoichiometries were determined by ICP-MS.

**Raman spectroscopy, XPS and XAS characterization**

Raman spectroscopic investigations were conducted with a WITec Raman system, operated with a green solid-state laser ($\lambda$=532 nm), a ×50 objective lens, a 1800 l/mm diffraction grating, and a Peltier-cooled charge-coupled device (CCD) detector. The Raman spectra were collected for 2 s with 10 accumulations each, with an incident laser power of ca. 0.02 mW on the sample surface to avoid any potential laser-induced oxidation[51].

The electrodes were studied *ex-situ* before and after 27 h of electrolyte contact at OCP by XPS using monochromated Al K$\alpha$ 1486.6 eV radiation. Spectra were recorded with a pass energy of 55 eV and a step size of 0.2 eV. Binding energies were calibrated by positioning the main graphite C 1s peak at 284.5 eV.

All the presented XANES spectra at Mo K- and Re L-edges were collected in fluorescence mode using the beam line 10C of the Pohang Accelerator Laboratory in Korea. Reference spectra were simultaneously measured using Mo and Re metal foils for energy calibration.

**Electron microscopy characterization**

SEM measurements were performed on a Hitachi S-4800 at 15 kV and a Zeiss Gemini at 1.5 kV. HRTEM experiments were conducted at 300 kV on a Titan Themis (Thermo Fisher) microscope equipped with an aberration corrector for the objective lenses. STEM measurements were done at 300 kV on a Titan Themis microscope (Thermo Fisher) equipped with an aberration corrector for the condenser lenses. EELS and EDS measurements were recorded using a Quantum ERS spectrometer (Gatan) and a SuperX detector (Thermo Fisher), respectively.

**Electrochemical characterization**

The $Re_xMo_{1-x}S_2$/CP electrodes were electrochemically characterized using a Metrohm Autolab PGSTAT204 potentiostat with a reversible hydrogen electrode (RHE, Hydroflex®) as the reference and glassy carbon as the counter electrode. The electrodes were stabilized for 32 h at OCP and the electrolyte was changed before starting HER. Such pretreatment was introduced to decouple the contributions of first electrolyte contact. If such contributions are not decoupled, a decay in the HER response after 4000 CV could be mistakenly attributed to HER corrosion, even if it is instead caused by the first electrolyte contact dissolution. HER stability was evaluated by 4000 cycles of CV at 100 mV/s of scan rate in a 0.5 M $H_2SO_4$ electrolyte saturated with Ar gas through continuous purging. Three cycles were recorded at a scan rate of 1 mV/s every 1000 cycles. Electrochemical impedance spectroscopy was also acquired at –100 $mV_{RHE}$ from $10^5$ to 10 Hz to apply 100% IR correction to all CV scans during data analysis. Chronopotentiometry measurements were conducted at 0 A for 20 h to evaluate the OCP of the electrodes using glassy carbon as counter electrode and RHE as reference electrode. To understand the effect of OCP on the HER activity, pretreated electrodes with 32 h of OCP were measured for HER for 3 cycles at scan rate of 1 mV/s at this initial state (Day 0) and after 13 (Day 13) and 30 (Day 30) days of electrolyte contact at OCP. The electrolyte (5 ml) was exchanged after the 13 days measurement.

To perform ILSTEM, $MoS_2$ and $Re_{0.2}Mo_{0.8}S_2$ nanocatalysts were drop cast on Au/C finder grids (Plano). To measure OCP, the grids were connected to the working electrode of a Gamry Reference 600 potentiostat through an Au wire (>99.99%, Redoxme) inserted in the grid. A RHE and a glassy carbon electrode were used as the reference and counter electrodes, respectively. The grids were left in contact with the electrolyte and were analyzed in the microscope fresh and after 13 days of electrolyte exposure. ILSEM measurements were conducted in cross section of the edge of the electrodes before and after 32 h of contact with the electrolyte.

For SFC-ICPMS measurements[52] a Gamry Reference 600 potentiostat, a saturated Ag/AgCl reference electrode (Metrohm) and a Pt wire counter electrode (99.997%, Alfa Aesar) were used. The 0.1 M $H_2SO_4$ electrolyte was analyzed online with a NexION300X spectrometer for Mo and Re elements. Y and Ir were added to the electrolyte in a 0.1 M $H_2SO_4$ solution as they served as internal standards to quantify Mo and Re, respectively.

**Computational methods**

Spin-polarized density functional theory calculations were performed using the Vienna Ab initio Simulation Package and employed the projected-augmented wave method[53, 54]. The exchange-correlation interaction was described using the strongly constrained and appropriately normed (SCAN) functional[55]. The DFT+U method was adopted applying a Hubbard U parameter of 2.05 eV for Mo[56]. The plane wave energy cutoff was set to 520 eV. The convergence criteria for the electronic total energy and the atomic force on atoms during all structural relaxations were set to $10^{-5}$ eV and 0.02 eV/Å, respectively. The Brillouin zone was sampled with a *k*-point grid density of 100 per reciprocal space. All crystal structures for the chemistries under study were sourced from the Materials Project[57] and optimized using the SCAN functional. The energetics obtained were employed to construct the Pourbaix diagram, utilizing a modified version of the method used in the Materials Project[56, 58]. The stable domain in the Pourbaix diagram is identified by computing all possible equilibrium redox reactions in an aqueous solution. These reactions are represented by the equation: [reactants] + $H_2O$ ↔ [products] + $mH^+$ + $ne^-$. The Nernst equation connects the Gibbs free energy change of these reactions to the surrounding pH and external potentials. Quantitatively, the aqueous stability of a catalyst is determined by comparing its chemical potential to that of the stable domains on the Pourbaix diagram under operating conditions.

**Supporting Information**

Supporting Information is available from the Wiley Online Library or from the author.

**Acknowledgements**

Petra Ebbinghaus is acknowledged for the initial Raman measurements. Benjamin Breitbach is acknowledged for the XRD data. Bettina V. Lotsch is acknowledged for providing the infrastructure for electrochemical measurements. R. Aymerich-Armengol is grateful for financial support from the International Max Planck Research School for Interface Controlled Materials for Energy Conversion (IMPRS-SurMat). M. Rabe's work is funded by the Deutsche Forschungsgemeinschaft under Germany's Excellence Strategy - EXC 2033 - 390677874 – RESOLV. F. Podjaski gratefully acknowledges UKRI funding under the grant reference EP/X027449/1. This research was supported by Nano·Material Technology Development Program through the National Research Foundation of Korea (NRF) funded by Ministry of Science and ICT (NRF-2022M3H4A4097520). We acknowledge support of the National Research Foundation of Korea (NRF) with a grant funded by the Korean government (MSIT) (No. NRF-2021R1F1A1061943). This research was supported by Learning & Academic research institution for Master's, PhD students, and Postdocs (LAMP) Program (No. RS-2023-00301850) and Basic Science Research Program (RS-2023-00271205) of NRF grant funded by the Korean Ministry of Education. The support under the framework of international cooperation program managed by the National Research Foundation of Korea (NRF-2023K2A9A2A22000124) and the Deutsche Forschungsgemeinschaft (DFG: ZH1105/3-1) is gratefully acknowledged.

**Operando Insights on the Degradation Mechanisms of Rhenium-doped and Undoped Molybdenum Disulfide Nanocatalysts for Electrolyzer Applications**

R. Aymerich-Armengol*, M. Vega-Paredes, Z. Wang, A. M. Mingers, L. Camuti, J. Kim, J. Bae, I. Efthimiopoulos, R. Sahu, F. Podjaski, M. Rabe, C. Scheu, J. Lim* and S. Zhang*

The degradation mechanisms of $Re_xMo_{1-x}S_2$ nanomaterials are described during HER and OCP by *operando* techniques. The electrochemical changes are correlated to the structural evolution of the nanocatalysts by electron microscopy and spectroscopy performed at identical locations. Rhenium-doped $MoS_2$ shows higher activity and stability than $MoS_2$ during HER.

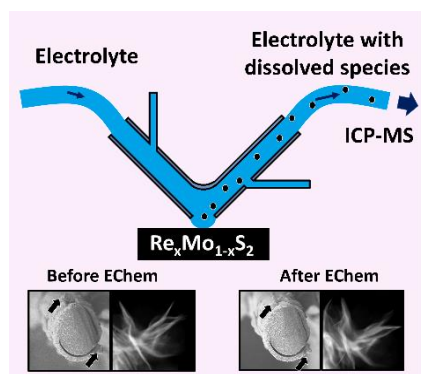